*(IJCSIS) International Journal of Computer Science and Information Security,*
*Vol. 6, No. 3, 2009*
# HIGH-PRECISION HALF-WAVE RECTIFIER CIRCUIT IN DUAL PHASE OUTPUT MODE

Theerayut Jamjaem
Department of Electrical Engineering
Faculty of Engineering, Kasem Bundit University
Bangkok, Thailand 10250
.

Bancha Burapattanasiri
Dept. of Electronic and Telecommunication Engineering
Faculty of Engineering, Kasem Bundit University
Bangkok, Thailand 10250
.
*Abstract*—This paper present high-precision half-wave rectifier circuit in dual phase output mode by 0.5 μm CMOS technology, ± 1.5 V low voltage, it has received input signal and sent output current signal, respond in high frequency. The main structure compound with CMOS inverter circuit, common source circuit, and current mirror circuit. Simulation and confirmation quality of working by PSpice program, then it able to operating at maximum frequency about 100 MHz, maximum input current range about 400 μA$_{p-p}$, high precision output signal, low power dissipation, and uses a little transistor.

## I. INTRODUCTION

Rectifier circuit is important circuit in analog working such as AC meter, detection signal circuit, and analog adaptation working circuit etc. And then it always has development and designing about rectifier circuit voltage mode. At the beginning used vacuum tube, diode [1-3] and next time used bipolar transistor [4]. At the beginning of rectifier circuit by diode and bipolar transistor, it has zero-crossing signal error and low precision. Because of diode and bipolar transistor have to use voltage driver working power around 0.3 V for Ge and 0.7 V for Si. So, the lower signal circuit unable to working. Next time, the limited of lower signal circuit has development such as rectifier circuit by Op-amp connected to diode, rectifier circuit by Op-amp connected to diode and bipolar transistor, and rectifier circuit AB class mode [9]. The result from circuit development is able to rectify the limited of lower signal circuit [1-5], responsiveness at narrowness frequency [2-7] and high dissipation current source to transistor. So, this paper is present a new choice of easy to understand and noncomplex of rectifier circuit, but it still have high quality in working function, it able working at high frequency, responding at high input current, high precision signal, low power dissipation, and uses a little transistor.

## II. DESIGNATION AND FUNCTIONAL

Three part component of the high-precision half-wave rectifier circuit in dual phrase output mode that is CMOS inverter circuit, it has functional to comparison positive part signal and negative part signal. Common source circuit, it has functional to sending positive part signal to current mirror circuit, and current mirror circuit, it has functional to establish half-wave positive and negative phase signal. For easy to understand, see diagram block in figure 1. The functional of circuit started when sending input positive and negative part signal to CMOS inverter circuit for comparison between negative ground part signal and positive to common source circuit and part to sending current mirror circuit 1, it has functional to establish half-wave positive phase signal. At that time positive part signal from common source circuit, it will send to current mirror circuit 2, it has functional to establish half-wave negative phase signal.

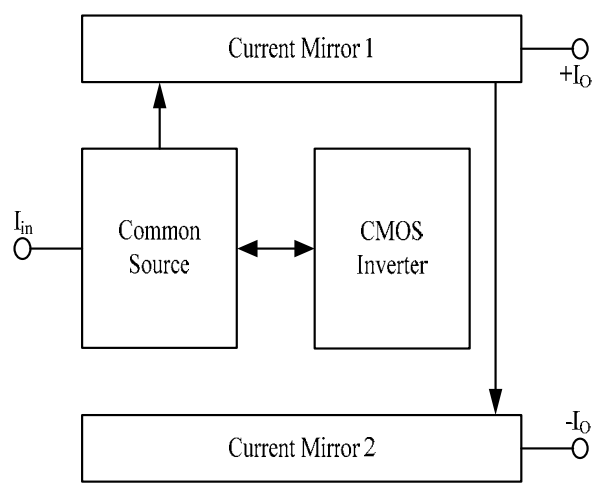

Figure 1. Diagram is show high-precision half-wave rectifier circuit in dual phase output mode.

From figure 1 is able to bring three part of circuit put together after that there are complete high-precision half-wave rectifier circuit in dual phase output mode as figure 2 and when you see from the structure of circuit by sending input current signal, it has equivalent to little than zero as an algebraic equation (1) after that is reflect to $M_3$ and $M_4$ transistor stop working, output voltage return to input of $M_1$ transistor common source circuit current the result is $I_{DM1}$ equivalent to

Identify applicable sponsor/s here. *(sponsors)*149

http://sites.google.com/site/ijcsis/
ISSN 1947-5500

$I_{in}$, but in directly opposite input current ($I_{in}$) equivalent to more zero is reflect to $M_1,M_3$ transistor stop working but $M_2, M_4$ are working, so the result is $I_{DM1}$ equivalent to zero as an algebraic equation (2).

$$I_{DM2} = 0 \text{ And } I_{DM1} = I_{in} \text{ When } I_{in} < 0 \quad (1)$$

$$I_{DM1} = 0 \text{ When } I_{in} > 0 \quad (2)$$

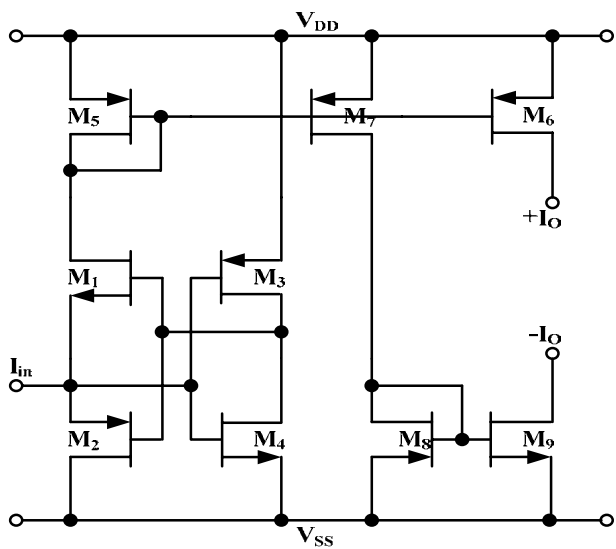

Figure 2. Completely high-precision half-wave rectifier circuit in dual phase output mode.

While $M_1$ transistor current is working, the signal current it has equivalent to little than zero is equivalent current it has $M_5$ transistor at drain pin. It has reflected current pass to transistor $M_6$, so the characteristic of output signal is half-wave negative phase as an algebraic equation (3) at the same time signal current at drain pin of $M_7$ transistor it has passed to drain pin of $M_8$ transistor, so $M_8$ and $M_9$ transistor are bonding in mirror current. So the characteristic of output signal is half-wave positive phase as an algebraic equation (4).

$$I_{DM1} = I_{DM5} = I_{DM6} = -I_{out} \quad (3)$$

$$I_{DM7} = I_{DM8} = I_{DM9} = +I_{out} \quad (4)$$

### III. SIMULATION AND MEASUREMENT RESULT

For confirmation of circuit working function uses PSpice program in testing, so we have to fix MIETEC parameter 0.5 µm for PMOS and NMOS transistor, power supply $V_{DD}$ = 1.5 V, $V_{SS}$ = -1.5 V, by sending input current signal at 400 µ$A_{p-p}$, frequency start from 1 kHz – 100 MHz and to setting W/L = 1.5/0.15 µm, at input current signal at 400 µ$A_{p-p}$, frequency 1 kHz, then the result is output signal as figure 3, at input current signal at 400 µ$A_{p-p}$, frequency 10 kHz, then the result is output signal as figure 4, at input current signal at 400 µ$A_{p-p}$, frequency 100 kHz, then the result is output signal as figure 5, at input current signal at 400 µ$A_{p-p}$, frequency 1 MHz, then the result is output signal as figure 6, at input current signal at 400 µ$A_{p-p}$, frequency 10 MHz, then the result is output signal as figure 7, at input current signal at 400 µ$A_{p-p}$, frequency 100 MHz, then the result is output signal as figure 8, Output signal at input current signal at 400 µ$A_{p-p}$, frequency 10 MHz, temperature 25°, 50°,75°and 100$^O$ as figure 9 and characteristic DC current at 400 µ$A_{p-p}$ input current, temperature 25°, 50°,75°and 100$^O$ as figure 10.

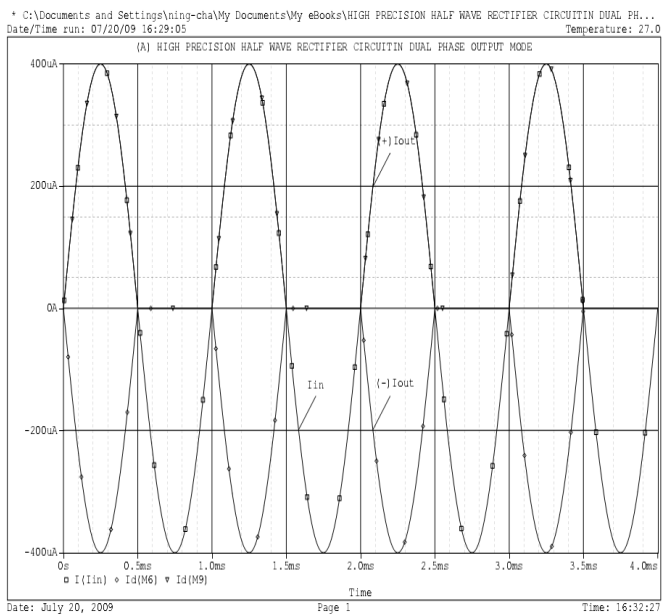

Figure 3. Output signal at input current 400 µ$A_{p-p}$ and frequency 1 kHz

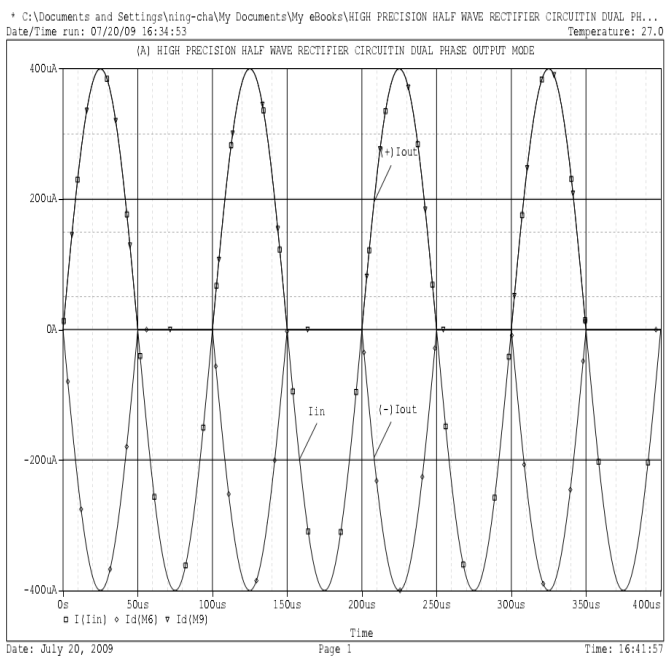

Figure 4. Output signal at input current 400 µ$A_{p-p}$ and frequency 10 kHz

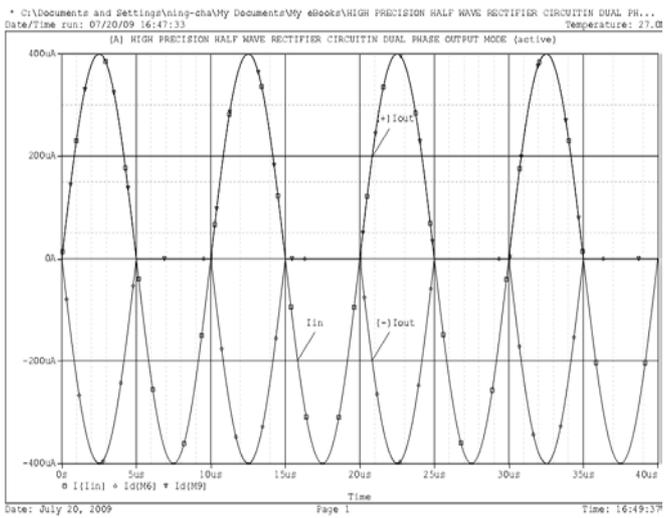

Figure 5. Output signal at input current 400 $\mu A_{p-p}$ and frequency 100 kHz

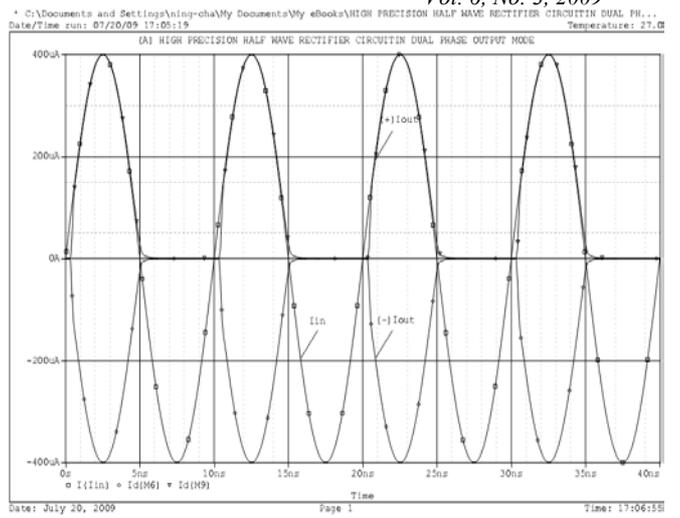

Figure 8. Output signal at input current 400 $\mu A_{p-p}$ and frequency 100MHz

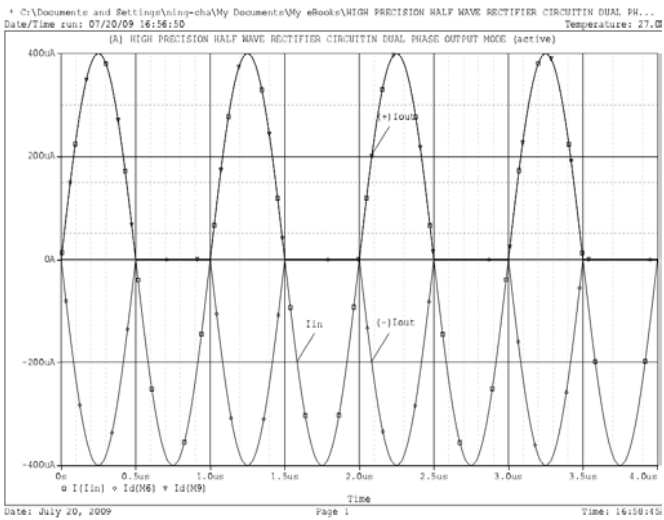

Figure 6. Output signal at input current 400 $\mu A_{p-p}$ and frequency 1 MHz

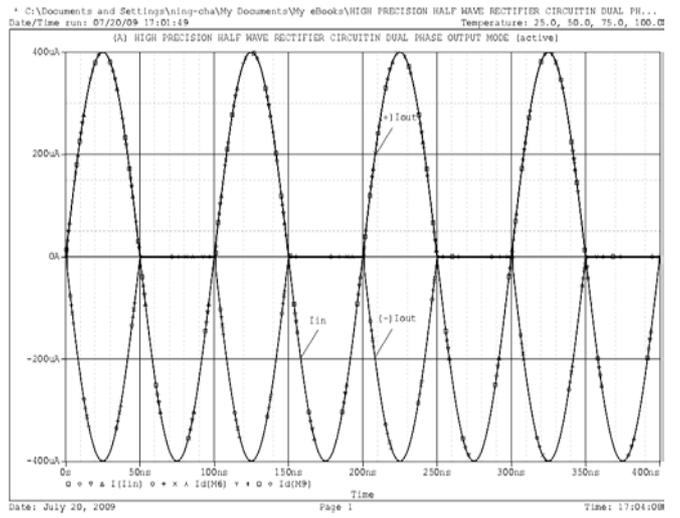

Figure 9. Output signal at input current 400 $\mu A_{p-p}$ and frequency 10 MHz at temperature $25^O$, $50^O$, $75^O$ and $100^O$

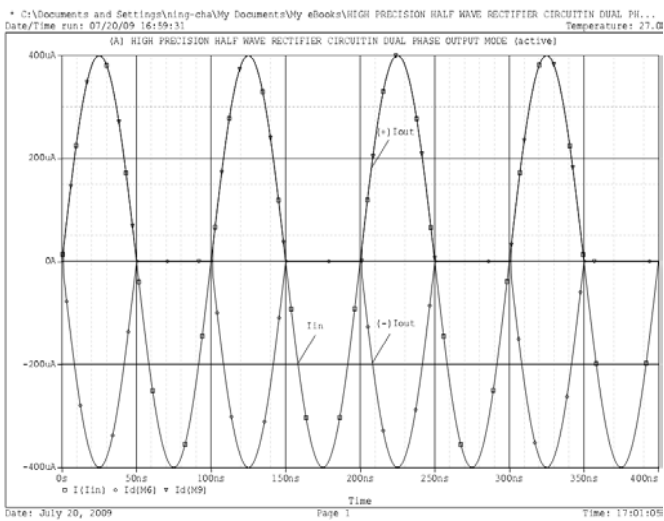

Figure 7. Output signal at input current 400 $\mu A_{p-p}$ and frequency 10 MHz

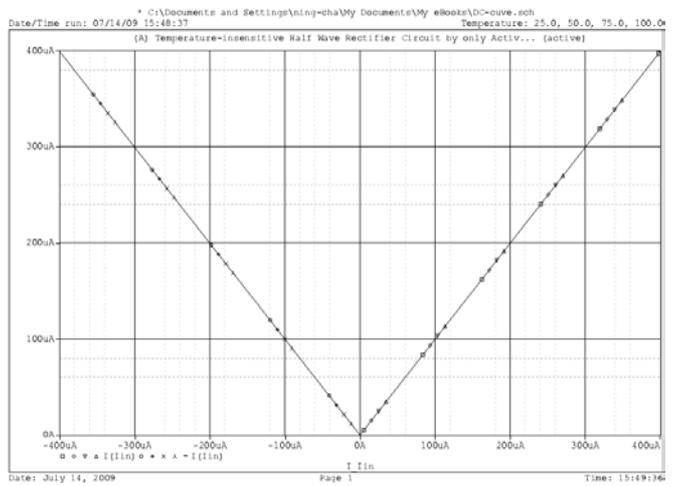

Figure 10. Output characteristic DC current at 400$\mu A_{p-p}$ input current, temperature $25^O$, $50^O$, $75^O$ and $100^O$

## IV. CONCLUSION

This present is show the circuit, it has component with a little of transistor, noncomplex in working function, dissipation of current source, working at input current mode, output signal is half-wave rectifier circuit in dual phase, without the reflection of temperature with $\pm$ 1.5 V low voltage and it not chance the structure of circuit. The result of testing it able to guarantee a quality of working function at maximum frequency 100 MHz, maximum output current 400 $\mu A_{p\text{-}p}$, losses power 198 pW. So, it suitable to uses in development VLSI compound current technology and apply in analog signal processing.

## APPENDIX

The parameters used in simulation are 0.5 µm CMOS Model obtained through MIETEC [10] as listed in Table I. For aspect ratio (W/L) of MOS transistors used are as follows: 1.5 µm / 0.15 µm for all NMOS transistors; 1.5 µm / 0.15 µm for all PMOS transistors.

TABLE I.  CMOS MODEL USED IN THE SIMULATION

```
-------------------------------------------------------------------------------------
.MODEL CMOSN NMOS LEVEL = 3 TOX = 1.4E-8 NSUB = 1E17
GAMMA = 0.5483559 PHI = 0.7 VTO = 0.7640855 DELTA = 3.0541177
UO = 662.6984452 ETA = 3.162045E-6 THETA = 0.1013999
KP = 1.259355E-4 VMAX = 1.442228E5 KAPPA = 0.3 RSH = 7.513418E-3
NFS = 1E12 TPG = 1 XJ = 3E-7 LD = 1E-13 WD = 2.334779E-7
CGDO = 2.15E-10 CGSO = 2.15E-10 CGBO = 1E-10 CJ = 4.258447E-4
PB = 0.9140376 MJ = 0.435903 CJSW = 3.147465E-10 MJSW = 0.1977689

.MODEL CMOSP PMOS LEVEL = 3 TOX = 1.4E-8 NSUB = 1E17
GAMMA = 0.6243261 PHI = 0.7 VTO = -0.9444911 DELTA = 0.1118368
UO = 250 ETA = 0 THETA = 0.1633973 KP = 3.924644E-5 VMAX = 1E6
KAPPA = 30.1015109 RSH = 33.9672594 NFS = 1E12 TPG = -1 XJ = 2E-7
LD = 5E-13 WD = 4.11531E-7 CGDO = 2.34E-10 CGSO = 2.34E-10
CGBO = 1E-10 CJ = 7.285722E-4 PB = 0.96443 MJ = 0.5
CJSW = 2.955161E-10 MJSW = 0.3184873
-------------------------------------------------------------------------------------
```

## ACKNOWLEDGMENT

The researchers, we are thank you very much to our parents, who has supporting everything to us. Thankfully to all of professor for knowledge and a consultant, thank you to Miss Suphansa Kansa-Ard for her time and supporting to this research. The last one we couldn't forget that is Kasem Bundit University, Engineering Faculty for supporting and give opportunity to our to development in knowledge and research, so we are special thanks for everything.

## REFERENCES

[1] Barker, R.W.J., "Versatile precision full wave Rectifier," Electron Letts, 5: Vol.13, pp. 143-144, 1977.
[2] Barker, R.W.J. and Hart, B.L., "Precision absolute-value circuit technique," Int. J. Electronics Letts, 3: Vol.66, pp. 445-448, 1989.
[3] Toumazou, C. and Lidgey, F.J., "Wide-Band precision rectification," IEE Proc. G, 1: Vol.134, pp.7-15, 1987.
[4] Wang, Z., "Full-wave precision rectification that is performed in current domain and very suitable for CMOS implementation," IEEE Trans. Circuits and Syst, 6: Part I, Vol. 39, pp.456-462, 1992.
[5] Toumazou, C., Lidgey, F.J.and Chattong, S., "High frequency current conveyor precision full-wave rectifier," Electron. Letts, 10: Vol. 30, pp. 745-746, 1994.
[6] Wilson, B. and Mannama, V., "Current-mode rectifier with improved precision," Electron. Letts, 4: Vol. 31, pp. 247-248, 1995.
[7] Surakampontorn, W. and Riewruja, V., "Integrable CMOS sinusoidal frequency doubler and full-wave rectifier," Int.J. Electronics, Letts, 3: Vol. 73, pp. 627-632, 1992.
[8] Traff, H., "Novel Approach to High Speed CMOS Current Comparators," Electron. Letts, 3: Vol. 28, pp. 310-312, 1992.
[9] Monpapassorn, A., "Improved Class AB Full-Wave rectifier," Thammasat Int. J. Sc. Tech., No. 3, November, Vol. 4, 1999.
[10] A. Monpapassorn, K. Dejhan, and F.Cheevasuvit, "CMOS dual output current mode half-wave rectifier," International Journal of Electronics, Vol. 88, 2001, pp. 1073-1084.

## AUTHORS PROFILE

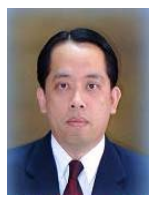

Mr.Theerayut Janjaem received the master degree in Telecommunication Engineering, from King Mongkut's Institute of Technology Ladkrabang in 2005. He is a lecture of Electrical Engineering Faculty of Engineering, Kasem Bundit University, Bangkok, Thailand. His research interests are energy, analog circuit design, low voltage, high frequency and high-speed CMOS technology.

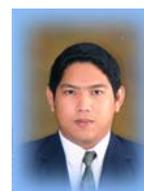

Mr.Bancha Burapattanasiri received the bleacher degree in electronic engineering from Kasem Bundit University in 2002 and master degree in Telecommunication Engineering, from King Mongkut's Institute of Technology Ladkrabang in 2008. He is a lecture of Electronic and Telecommunication Engineering, Faculty of Engineering, Kasem Bundit University, Bangkok, Thailand. His research interests analog circuit design, low voltage, high frequency and high-speed CMOS technology.